\documentclass[submitting]{nst}

\usepackage{subfigure,dcolumn}
\usepackage[T2A,T1]{fontenc}
\usepackage[russian,english]{babel}

\usepackage{listings}
\usepackage{float}
\newcommand{\snn}{\sqrt{s_\text{NN}}}

\begin{document}

\title{Effect of Light Nuclei on Chemical Freeze-out Parameters at RHIC Energies}\thanks{This work was supported by the Scientific Research Foundation of Hubei University of Education for Talent Introduction (Nos. ESRC20230002 and ESRC20230007) and the Research Project of Hubei Provincial Department of Education (Nos. D20233003 and B2023191).}

\author{Ning Yu}
\email[Corresponding author,]{Ning Yu, ning.yuchina@gmail.com}
\author{Zuman Zhang}
\author{Hongge Xu}
\affiliation{School of Physics and Mechanical Electrical and Engineering, Hubei University of Education, Wuhan 430205, China}
\affiliation{Institute of Astronomy and High Energy Physics, Hubei University of Education, Wuhan 430205, China}
\author{Minxuan Song}
\affiliation{School of Physics and Mechanical Electrical and Engineering, Hubei University of Education, Wuhan 430205, China}

\begin{abstract}
  In this study, the chemical freeze-out of hadrons, including light-and strange-flavor particles and light nuclei, produced in Au+Au collisions at the Relativistic Heavy Ion Collider (RHIC), was investigated. Using the thermal-FIST thermodynamic statistical model, we analyzed various particle sets: those inclusive of light nuclei, those exclusive to light nuclei, and those solely comprising light nuclei. We determined the chemical freeze-out parameters at $\snn=$ 7.7--200 GeV and four different centralities. A significant finding was the decrease in the chemical freeze-out temperature $T_{\textrm{ch}}$ with light nuclei inclusion, with an even more pronounced reduction when considering light nuclei yields exclusively. This suggests that light nuclei formation occurs at a later stage in the system's evolution at RHIC energies. We present parameterized formulas that describe the energy dependence of $T_{\textrm{ch}}$ and the baryon chemical potential $\mu_B$ for three distinct particle sets in central Au+Au collisions at RHIC energies. Our results reveal at least three distinct $T_{\textrm{ch}}$ at RHIC energies correspond to different freeze-out hypersurfaces: a light-flavor freeze-out temperature of $T_L$ = 150.2$\pm$6 MeV, a strange-flavor freeze-out temperature $T_s$ = 165.1$\pm$2.7 MeV, and a light-nuclei freeze-out temperature $T_{\textrm{ln}}$ = 141.7$\pm$1.4 MeV. Notably, at the Large Hadron Collider (LHC) Pb+Pb 2.76 TeV, the expected lower freeze-out temperature for light nuclei was not observed; instead, the $T_{\textrm{ch}}$ for light nuclei was found to be approximately 10 MeV higher than that for light-flavor hadrons.
\end{abstract}
  

\maketitle

\section{Introduction}

Quantum chromodynamics (QCD) is the fundamental theory of strong interactions that governs the strong interactions between quarks and gluons and serves as the cornerstone for understanding QCD matter. One of the major goals of ultra-relativistic nuclear collisions is to explore the properties of the QCD phase diagram~\cite{RN168}, which shows the possible phases of QCD matter under different temperatures and baryon density conditions. According to lattice QCD calculations, a deconfinement transition from hadronic matter to a new state of matter called quark-gluon plasma (QGP) has been predicted at high temperatures and low baryon densities ~\cite{RN47}. QGP is a state in which quarks and gluons are no longer confined inside hadrons but can move freely in a hot and dense medium. The existence of QGP has been confirmed by various experimental signatures observed in heavy-ion collisions at ultra-relativistic energies, such as the suppression of high transverse momentum ($p_T$) hadrons owing to jet quenching~\cite{RN149,RN67} and the large elliptic flow ($v_2$) for hadrons owing to collective expansion~\cite{RN71,RN15,RN52,RN328,RN329,RN43}.

After the discovery of strongly coupled QGP, efforts have been made to vary the collision energy and explore the phase structure of hot and dense QCD matter, which can be represented by the $T-\mu_B$ plane ($T$: temperature, $\mu_B$: baryon chemical potential) of the QCD phase diagram. One of the most powerful tools for probing QCD phase diagrams is hadron production, which reflects the system’s thermodynamic conditions during a chemical freeze-out. Hadron yields were measured from the AGS to LHC energies and can be described by a thermal statistical model, assuming that chemical equilibrium is reached~\cite{RN103,RN24,RN41,RN102,RN327}. The temperature $T_{\textrm{ch}}$ and $\mu_B$ at chemical equilibrium, which are also referred to as the chemical freeze-out parameters when all hadron abundances are fixed, are determined from experimental data fitting.

One of the remarkable findings of previous studies on LHC energy was that the chemical freeze-out temperature $T_{\textrm{ch}}=$156.5$\pm$1.5 MeV obtained from the thermal fit to the hadron and light nuclei yields~\cite{RN102} was consistent with the pseudo-critical temperature $T_c = $154$\pm$9 MeV obtained from lattice QCD calculations~\cite{RN99} within a certain uncertainty. This suggests that chemical freeze-out occurred close to the phase boundary between the QGP and hadronic matter. The data used for the thermal fit were from ALICE Collaboration, which measured not only hadrons but also light nuclei, such as $d(\bar{d})$, $^3\textrm{He}(^3\overline{\textrm{He}})$, $^4\textrm{He}(^4\overline{\textrm{He}})$, and hyper-triton $^3_{\Lambda}\textrm{H}$~\cite{RN102}. Meanwhile, a $T_{\textrm{ch}} of $167.8$\pm$4.2 MeV extracted via thermal fitting at the RHIC top energy was higher than the value at the LHC and $T_c$~\cite{RN24}, while only hadron yields were included in the fitting. Additionally, the authors also reported that when the fit is limited to the yields of pions, kaons, and protons, the derived $T_{\textrm{ch}}$ is approximately 10--15 MeV lower than that of fits that include the yields of hadrons with strange quarks. Furthermore, at the LHC energy, it was found that considering two distinct freeze-out temperatures for non-strange and strange hadrons could provide a better fit to the ALICE data~\cite{RN307}. Specifically, by analyzing energies ranging from 11.5 GeV to 5.02 TeV in Ref. ~\cite{RN308}, a light-flavor freeze-out temperature of $T_L$=150.2$\pm$6 MeV and a strange-flavor freeze-out temperature of $T_s$=165.1$\pm$2.7 MeV at vanishing $\mu_B$ were identified. These findings indicate that strange and non-strange hadrons are produced at distinct freeze-out hypersurfaces, which shows evidence of flavor-dependent chemical freeze-out temperatures, at least in the crossover region of the QCD phase diagram. However, these results do not fully account for the differences in $T_{\textrm{ch}}$ between the RHIC top energy and LHC energy, as both fits include yields for both strange and non-strange hadrons. By comparing the particles used in the fits at RHIC and LHC, it appears that the yield of light nuclei may be the only factor that could account for the differences in the results. Therefore, this study aimed to investigate whether the inclusion of light nuclei yields in RHIC experiments leads to changes in $T_{\textrm{ch}}$, and whether light nuclei have their own distinct $T_{\textrm{ch}}$.

In relativistic heavy-ion collisions, the detailed mechanisms underlying the production of light nuclei are not fully understood~\cite{RN295,RN97,RN100,RN86,RN113,RN136,RN216,RN320,RN321,RN326,RN325}. A widely discussed theory suggests that light nuclei are formed by the coalescence of nucleons during the final stages of collisions ~\cite{RN82,RN87}. This scenario clearly implies that the production of light nuclei occurs later than that of other hadrons and, consequently, at a lower temperature than that at which other hadrons are formed. Additionally, the thermal model offers an alternative perspective~\cite{RN102,RN101,RN163} in which protons, neutrons, and light nuclei are considered to reach a state of chemical equilibrium. In this scenario, light nuclei and nucleons are presumed to be produced concurrently, suggesting that their freeze-out temperatures are identical or similar to those of light-flavor hadrons. One particular study highlighted in Ref. ~\cite{RN100} demonstrated that the yield of (anti-)deuterons remained relatively stable throughout the system's evolution, given that it was thermally initialized at the LHC energy. The (anti-)deuteron can be destroyed and created through the reaction $\pi d \leftrightarrow\pi pn$, indicating that the thermal model, despite its different foundational assumptions, may yield similar (anti-)deuteron yields as the coalescence model. Therefore, studying the chemical freeze-out temperature of light nuclei can provide valuable insights into the light nuclei production mechanisms, aiding our understanding of the evolution of fireballs produced in relativistic heavy-ion collisions.

To study the effects of light nuclei on $T_{\textrm{ch}}$, we utilized the statistical thermal model Thermal-FIST to analyze the yields of hadrons~\cite{RN24}, (anti-)deuterons ~\cite{RN97}, and tritons ~\cite{RN218} at RHIC energies. We compared the extracted freeze-out parameters across various particle yield sets and discussed the underlying physics of the production of strange, light-flavor, and light nuclei. In Section II, we provide a brief introduction to the employed Statistical Thermal Model and selected particle sets. Subsequently, in Section III, we present and discuss the results. Finally, Section IV summarizes our study and findings.

\section{Model and Particle Sets}

The chemical properties of bulk particle production can be treated within the framework of thermal statistical models using the Thermal-FIST package~\cite{RN95}. This package is designed for convenient general-purpose physics analysis and is well-suited for the family of hadron resonance gas (HRG) models. Notable features of this package include the treatment of fluctuations and correlations of conserved charges, the effects of probabilistic decay, chemical non-equilibrium, and the inclusion of van der Waals hadronic interactions. However, in our study, we opted for an ideal non-interacting gas of hadrons and resonances within a Grand Canonical Ensemble (GCE) for simplicity and focus on the fundamental aspects of particle production. In GCE, all conserved charges, such as the baryonic number $B$, electric charge $Q$, strangeness $S$, and charm $C$, are conserved on average. For the chemical equilibrium case, these average values are regulated by their corresponding chemical potentials, $\mu_B, \mu_S, \mu_Q, and \mu_C$. The chemical potential $\mu_i$ of hadron species $i$ is determined as
\begin{equation}
  \mu_i=B_i\mu_B+S_i\mu_S+Q_i\mu_Q+C_i\mu_C.
\end{equation}
The particle density of hadron species $i$ can be parameterized as
\begin{equation}
  n_i(T,\mu_i) = g_i \int{\frac{d^3p}{\left(2\pi\right)^3}\left[ \gamma_s^{-\left|S_i^{'}\right|}\exp \left( {\frac{{{E_i} - {\mu_i}}}{{{T_{ch}}}}} \right) + \eta_i\right]^{-1}},
\end{equation}\label{NoverV}
where $g_i$, $E_i=\sqrt{p^2+m_i^2}$, and $m_i$ are the spin degeneracy factor, energy, and mass of hadron species $i$, respectively,. $\eta_i$ is $+1$ for fermions, $-1$ for bosons, and $0$ for the Boltzmann approximation. The quantum statistic is considered only for mesons, that is, $\eta_i = -1$ for $\pi^\pm$ and $K^\pm$, whereas $\eta_i = 0$ for other particles. Finite resonance widths with a Breit--Wigner form mass distribution were considered in our fitting~\cite{RN220}. $\mu_Q$ and $\mu_S$ are determined in a unique manner to satisfy two conservation laws given by the initial conditions: the electric-to-baryon charge ratio of $Q/B = 0.4$ for Au+Au collisions and the vanishing net strangeness $S = 0$. $\gamma_s$ is the strangeness undersaturation factor that regulates the deviation from the chemical equilibrium of strange quarks. $S_i^{'}$ is the number of strange and anti-strange valence quarks in particle $i$. $\gamma_s=1$ corresponds to the chemical equilibrium for strange quarks. 

The hadron yield of $\pi^\pm$, $K^\pm$, $p(\bar{p})$~\cite{RN23,RN24,RN311,RN105}, $\Lambda(\bar{\Lambda})$, $\Xi^-(\bar{\Xi}^+)$\cite{RN136,RN309,RN310} and light nuclei yield of $d(\bar{d})$~\cite{RN97}, and $t$~\cite{RN218} for Au+Au collision at $\snn=$ 7.7--200 GeV were used in the analysis. The yields of $p(\bar{p})$ are inclusive and are not corrected for the feed-down contribution from weak decay. The Thermal-FIST package allows the assignment of separate decay chains to each input. In the model, the total yield $N_i$ of the $i$-th particle species is calculated as the sum of the primordial yield $N_i^*$ and the resonance decay contributions, as follows:
\begin{equation}
  N_i = N_i^* + \sum_j \textrm{B.R}_{ji}\times N_j,
\end{equation}
Where $\textrm{B.R}_{ji}$ denotes the branching ratio of the $j$-th to the $i$-th particle species. Therefore, it matched the specific feed-down corrections of a particular dataset. Four centralities, 0
Three different sets of particle yields are listed in Table~\ref{tab1}. The first set is the same as that in Ref. ~\cite{RN24}, which includes the yields of $\pi^\pm$, $K^\pm$, $p(\bar{p})$, $\Lambda(\bar{\Lambda})$, and $\Xi^-(\bar{\Xi}^+)$. The yields of light nuclei $d(\bar{d})$ and $t$ are included in the second particle set. In particle set III, we excluded all mesons and hadrons, meaning that only $d(\bar{d})$ and $t$ are included. The three particle sets are presented in Table~\ref{tab1}. Therefore, only four parameters were fitted. For particle sets I and II, the fitting process involves four parameters: $T_{\text{ch}}$, the baryon chemical potential $\mu_B$, the strangeness undersaturation factor $\gamma_s$, and the fireball radius $V$ (the fireball volume is $V=4/3\pi R^3$). However, for particle set III, which does not include strange particles, the strangeness undersaturation factor $\gamma_s$ is fixed at 1, resulting in only three parameters being fitted. Because there are only three particle yields for fitting in particle set III, a perfect fit with zero degrees of freedom is obtained; therefore, a $\chi^2$/dof for the fit cannot be obtained.
\renewcommand{\arraystretch}{1.2}
\begin{table}[!htb]
  \caption{Particle sets in the Thermal-FIST fit via GCE.}
  \label{tab1}
  \begin{tabular*}{\hsize}{@{\extracolsep{\fill}} cc}
  \toprule
  Particle Set	&	Particle list \\
  \cmidrule(r){1-2}
  I	  &	$\pi^\pm$, $K^\pm$, $p$($\bar{p}$), $\Lambda$($\bar{\Lambda}$), $\Xi^-$($\bar{\Xi}^+$)\\
  II	&	$\pi^\pm$, $K^\pm$, $p$($\bar{p}$), $\Lambda$($\bar{\Lambda}$), $\Xi^-$($\bar{\Xi}^+$), $d$($\bar{d}$), t\\
  III	&	$d$($\bar{d}$), t\\
  \bottomrule
  \end{tabular*}
\end{table}
\renewcommand{\arraystretch}{1}

\section{Results}

The detailed fitting outcomes for each energy, encompassing the fireball volume $V$ and $\chi^{2}$/dof, are delineated in Table~\ref{tab2} for collision energies 
$\snn=$ 200, 62.4, 39, and 27 GeV. For the lower collision energies of $\snn=$ 19.6, 14.5, 11.5, and 7.7 GeV, these are presented in Table~\ref{tab3}. It is noteworthy that for the particle set III, at 11.5 GeV of 40-80\% centrality, and 7.7 GeV of all centralities, the experimental data for $\bar{d}$ is absent. Due to the lack of sufficient experimental data, specifically with only two yield data points ($d$ and $t$) for the three fitting parameters ($T_{\text{ch}}$, $\mu_B$, $V$), it is impossible to accurately determine the chemical freeze-out parameters.

\begin{table}[!htb]
  \caption{The Thermal-FIST Grand Canonical Ensemble fit for collision energies $\snn$ = 200, 62.4, 39, 27 Gev. For the fits in particle set I and II, $T_{\text{ch}}$, $\mu_B$, $\gamma_s$, and $V$ were used as free parameters. For the fit in particle set III, $\gamma_s$ were fixed to unity.}
  \label{tab2}
  \footnotesize
  \begin{tabular*}{\hsize} {@{\extracolsep{\fill} } lccccc}
  \toprule
  Centrality&$T_{\text{ch}}$(\si{MeV}) & $\mu_B$(\si{MeV}) & $\gamma_s$ & $V$(\si{fm^3}) & $\chi^2$/dof \\
  \midrule
  \multicolumn{6}{c}{$\snn=$ 200 GeV Pariticle Set I} \\
  0-10\% &163.5$\pm$3.8&23.3$\pm$9.2&1.09$\pm$0.06&1389$\pm$256&11.2/6\\
  10-20\%&161.3$\pm$3.4&22.0$\pm$8.0&1.03$\pm$0.05&1028$\pm$172&13.0/6\\
  20-40\%&163.3$\pm$3.2&21.3$\pm$6.2&1.02$\pm$0.04&542$\pm$85&17.0/6\\
  40-80\%&162.7$\pm$3.3&17.3$\pm$7.7&0.92$\pm$0.04&164$\pm$5&14.7/6\\
  \midrule
  \multicolumn{6}{c}{$\snn=$ 200 GeV Pariticle Set II} \\
  0-10\% &152.0$\pm$1.1&18.1$\pm$3.1&1.23$\pm$0.05&2436$\pm$217&39.7/9\\
  10-20\%&154.2$\pm$1.1&18.0$\pm$3.0&1.10$\pm$0.03&1441$\pm$129&33.9/9\\
  20-40\%&154.9$\pm$1.0&16.3$\pm$2.8&1.12$\pm$0.02&800$\pm$65&40.9/9\\
  40-80\%&153.5$\pm$1.1&15.3$\pm$2.9&1.01$\pm$0.03&247$\pm$21&34.9/9\\
  \midrule
  \multicolumn{6}{c}{$\snn=$ 200 GeV Pariticle Set III} \\
  0-10\% &137.6$\pm$3.4&26.1$\pm$3.2&1&11210$\pm$4410&$\backslash$\\
  10-20\%&139.6$\pm$3.9&25.1$\pm$3.2&1&6650$\pm$2904&$\backslash$\\
  20-40\%&141.6$\pm$3.3&23.3$\pm$3.2&1&3214$\pm$1158&$\backslash$\\
  40-80\%&142.3$\pm$3.8&20.4$\pm$3.2&1&771$\pm$314&$\backslash$\\
  \midrule
  \multicolumn{6}{c}{$\snn=$ 62.4 GeV Pariticle Set I} \\
  0-10\% &159.3$\pm$3.1&66.2$\pm$10.4&0.99$\pm$0.05&1220$\pm$187&16.2/6\\
  10-20\%&158.5$\pm$2.7&64.3$\pm$8.8&1.00$\pm$0.05&842$\pm$120&16.4/6\\
  20-40\%&158.8$\pm$2.7&58.6$\pm$8.5&0.99$\pm$0.05&484$\pm$68&14.7/6\\
  40-80\%&157.5$\pm$2.6&52.1$\pm$7.7&0.89$\pm$0.05&143$\pm$21&17.6/6\\
  \midrule
  \multicolumn{6}{c}{$\snn=$ 62.4 GeV Pariticle Set II} \\
  0-10\% &152.8$\pm$1.1&63.1$\pm$3.1&1.02$\pm$0.05&1669$\pm$151&34.0/9\\
  10-20\%&154.2$\pm$1.1&59.9$\pm$3.0&1.01$\pm$0.04&1057$\pm$96&30.9/9\\
  20-40\%&155.6$\pm$1.1&57.1$\pm$3.0&0.99$\pm$0.04&572$\pm$51&23.1/9\\
  40-80\%&155.4$\pm$1.4&53.8$\pm$3.2&0.90$\pm$0.04&158$\pm$17&19.6/9\\
  \midrule
  \multicolumn{6}{c}{$\snn=$ 62.4 GeV Pariticle Set III} \\
  0-10\% &140.6$\pm$3.2&66.9$\pm$3.0&1&6141$\pm$2233&$\backslash$\\
  10-20\%&142.8$\pm$3.1&63.5$\pm$3.0&1&3514$\pm$1218&$\backslash$\\
  20-40\%&146.5$\pm$3.3&60.3$\pm$3.1&1&1449$\pm$509&$\backslash$\\
  40-80\%&149.4$\pm$6.0&53.6$\pm$3.6&1&272$\pm$157&$\backslash$\\
  \midrule
  \multicolumn{6}{c}{$\snn=$ 39 GeV Pariticle Set I} \\
  0-10\% &157.2$\pm$3.1&104.0$\pm$9.3&1.12$\pm$0.07&954$\pm$168&6.5/6\\
  10-20\%&158.4$\pm$3.0&101.8$\pm$8.4&1.10$\pm$0.06&635$\pm$106&6.8/6\\
  20-40\%&160.4$\pm$3.1&94.6$\pm$7.7&1.02$\pm$0.05&346$\pm$57&6.2/6\\
  40-80\%&159.5$\pm$3.0&77.6$\pm$7.5&0.86$\pm$0.04&105$\pm$17&5.4/6\\
  \midrule
  \multicolumn{6}{c}{$\snn=$ 39 GeV Pariticle Set II} \\
  0-10\% &154.6$\pm$1.4&99.3$\pm$3.1&1.09$\pm$0.06&1163$\pm$136&20.1/9\\
  10-20\%&155.6$\pm$1.3&95.2$\pm$3.0&1.08$\pm$0.05&776$\pm$84&21.5/9\\
  20-40\%&156.1$\pm$1.2&87.3$\pm$2.8&1.03$\pm$0.03&448$\pm$44&22.6/9\\
  40-80\%&154.2$\pm$1.1&74.3$\pm$2.6&0.89$\pm$0.03&139$\pm$13&19.1/9\\
  \midrule
  \multicolumn{6}{c}{$\snn=$ 39 GeV Pariticle Set III} \\
  0-10\% &142.6$\pm$3.3&99.6$\pm$3.2&1&4179$\pm$1524&$\backslash$\\
  10-20\%&143.8$\pm$3.1&96.0$\pm$3.1&1&6650$\pm$2904&$\backslash$\\
  20-40\%&145.0$\pm$2.7&89.9$\pm$2.9&1&1448$\pm$431&$\backslash$\\
  40-80\%&145.5$\pm$2.5&78.2$\pm$2.8&1&350$\pm$96&$\backslash$\\
  \midrule
  \multicolumn{6}{c}{$\snn=$ 27 GeV Pariticle Set I} \\
  0-10\% &158.1$\pm$2.7&149.2$\pm$7.5&1.16$\pm$0.05&812$\pm$128&8.5/6\\
  10-20\%&158.6$\pm$2.6&142.6$\pm$7.0&1.11$\pm$0.05&573$\pm$91&8.2/6\\
  20-40\%&161.3$\pm$3.0&135.1$\pm$7.0&1.01$\pm$0.04&303$\pm$51&9.7/6\\
  40-80\%&163.1$\pm$3.0&113.6$\pm$6.7&0.82$\pm$0.03&83$\pm$14&7.3/6\\
  \midrule
  \multicolumn{6}{c}{$\snn=$ 27 GeV Pariticle Set II} \\
  0-10\% &154.3$\pm$1.1&140.1$\pm$3.0&1.17$\pm$0.04&1030$\pm$94&23.5/9\\
  10-20\%&154.6$\pm$1.1&134.3$\pm$3.0&1.12$\pm$0.03&733$\pm$66&25.6/9\\
  20-40\%&156.5$\pm$1.1&126.0$\pm$2.7&1.03$\pm$0.03&402$\pm$35&22.7/9\\
  40-80\%&155.2$\pm$1.0&103.9$\pm$2.7&0.88$\pm$0.02&123$\pm$10&26.0/9\\
  \midrule
  \multicolumn{6}{c}{$\snn=$ 27 GeV Pariticle Set III} \\
  0-10\% &144.0$\pm$2.8&138.0$\pm$4.0&1&3151$\pm$955&$\backslash$\\
  10-20\%&142.7$\pm$2.9&132.5$\pm$4.1&1&2656$\pm$857&$\backslash$\\
  20-40\%&146.9$\pm$2.9&125.9$\pm$4.0&1&1067$\pm$321&$\backslash$\\
  40-80\%&145.9$\pm$2.7&108.6$\pm$3.9&1&303$\pm$89&$\backslash$\\
  \bottomrule
  \end{tabular*}
\end{table}

\begin{table}[!htb]
  \caption{The Thermal-FIST Grand Canonical Ensemble fit for collision energies $\snn$ = 19.6, 14.5, 11.5, 7.7 Gev. For the fits in particle set I and II, $T_{\text{ch}}$, $\mu_B$, $\gamma_s$, and $V$ were used as free parameters. For the fit in particle set III, $\gamma_s$ were fixed to unity.}
  \label{tab3}
  \footnotesize
  \begin{tabular*}{\hsize} {@{\extracolsep{\fill} } lccccc}
  \toprule
  Centrality&$T_{\text{ch}}$(\si{MeV}) & $\mu_B$(\si{MeV}) & $\gamma_s$ & $V$(\si{fm^3}) & $\chi^2$/dof \\
  \midrule
  \multicolumn{6}{c}{$\snn=$ 19.6 GeV Pariticle Set I} \\
  0-10\% &156.0$\pm$2.7&194.4$\pm$8.5&1.15$\pm$0.05&801$\pm$131&7.7/6\\
  10-20\%&158.4$\pm$2.9&190.7$\pm$8.8&1.07$\pm$0.04&513$\pm$88&7.2/6\\
  20-40\%&159.6$\pm$2.8&177.6$\pm$8.2&0.99$\pm$0.04&542$\pm$85&6.2/6\\
  40-80\%&160.5$\pm$3.2&148.9$\pm$8.5&0.78$\pm$0.03&84.4$\pm$14.8&6.3/6\\
  \midrule
  \multicolumn{6}{c}{$\snn=$ 19.6 GeV Pariticle Set II} \\
  0-10\% &152.5$\pm$1.1&185.4$\pm$3.3&1.16$\pm$0.04&999$\pm$91&18.9/9\\
  10-20\%&153.1$\pm$1.1&178.6$\pm$3.3&1.10$\pm$0.04&698$\pm$65&23.6/9\\
  20-40\%&155.2$\pm$1.1&167.9$\pm$3.2&1.01$\pm$0.03&378$\pm$33&19.2/9\\
  40-80\%&155.0$\pm$1.2&141.7$\pm$3.2&0.82$\pm$0.03&111$\pm$10&16.6/9\\
  \midrule
  \multicolumn{6}{c}{$\snn=$ 19.6 GeV Pariticle Set III} \\
  0-10\% &143.3$\pm$2.9&182.1$\pm$5.5&1&2693$\pm$832&$\backslash$\\
  10-20\%&141.3$\pm$3.1&177.0$\pm$5.7&1&2353$\pm$792&$\backslash$\\
  20-40\%&144.4$\pm$3.3&165.8$\pm$5.7&1&1099$\pm$375&$\backslash$\\
  40-80\%&147.0$\pm$3.1&146.5$\pm$5.6&1&226$\pm$72&$\backslash$\\
  \midrule
  \multicolumn{6}{c}{$\snn=$ 14.5 GeV Pariticle Set I} \\
  0-10\% &153.4$\pm$3.1&243.7$\pm$13.2&1.05$\pm$0.06&818$\pm$149&6.4/6\\
  10-20\%&154.4$\pm$3.4&237.4$\pm$13.6&1.02$\pm$0.06&545$\pm$106&8.4/6\\
  20-40\%&155.4$\pm$3.2&225.2$\pm$12.7&0.94$\pm$0.05&309$\pm$55&5.3/6\\
  40-80\%&154.8$\pm$3.3&194.8$\pm$12.0&0.74$\pm$0.04&92.9$\pm$16.7&8.7/9\\
  \midrule
  \multicolumn{6}{c}{$\snn=$ 14.5 GeV Pariticle Set II} \\
  0-10\% &149.8$\pm$1.5&234.3$\pm$4.8&1.05$\pm$0.05&1026$\pm$114&19.1/9\\
  10-20\%&151.8$\pm$1.7&234.2$\pm$4.9&1.02$\pm$0.05&639$\pm$77&15.9/9\\
  20-40\%&152.4$\pm$1.5&220.3$\pm$4.3&0.95$\pm$0.05&369$\pm$40&12.8/9\\
  40-80\%&154.8$\pm$1.9&198.3$\pm$5.1&0.74$\pm$0.03&93.6$\pm$11.8&10.0/9\\
  \midrule
  \multicolumn{6}{c}{$\snn=$ 14.5 GeV Pariticle Set III} \\
  0-10\% &137.7$\pm$4.0&227.9$\pm$12.4&1&3593$\pm$1399&$\backslash$\\
  10-20\%&140.5$\pm$4.6&228.3$\pm$10.8&1&1888$\pm$889&$\backslash$\\
  20-40\%&143.6$\pm$3.5&219.8$\pm$9.3&1&849$\pm$285&$\backslash$\\
  40-80\%&148.3$\pm$6.6&194.7$\pm$12.2&1&167$\pm$103&$\backslash$\\
  \midrule
  \multicolumn{6}{c}{$\snn=$ 11.5 GeV Pariticle Set I} \\
  0-10\% &150.5$\pm$2.7&297.1$\pm$13.4&1.07$\pm$0.06&772$\pm$129&5.4/6\\
  10-20\%&152.1$\pm$2.7&297.0$\pm$12.5&1.06$\pm$0.05&475$\pm$80&6.9/6\\
  20-40\%&154.7$\pm$2.8&281.1$\pm$12.6&0.91$\pm$0.05&266$\pm$44&4.2/6\\
  40-80\%&157.5$\pm$3.2&254.8$\pm$13.2&0.71$\pm$0.04&68.3$\pm$12.3&6.0/6\\
  \midrule
  \multicolumn{6}{c}{$\snn=$ 11.5 GeV Pariticle Set II} \\
  0-10\% &147.2$\pm$1.3&281.3$\pm$4.5&1.07$\pm$0.05&975$\pm$96&18.1/9\\
  10-20\%&148.1$\pm$1.2&279.9$\pm$4.0&1.07$\pm$0.05&614$\pm$60&17.9/9\\
  20-40\%&149.0$\pm$1.2&266.2$\pm$3.8&0.96$\pm$0.04&362$\pm$33&21.8/9\\
  40-80\%&153.0$\pm$1.6&236.1$\pm$4.7&0.73$\pm$0.03&86.7$\pm$9.1&9.8/8\\
  \midrule
  \multicolumn{6}{c}{$\snn=$ 11.5 GeV Pariticle Set III} \\
  0-10\% &138.9$\pm$3.3&265.8$\pm$9.6&1&2771$\pm$903&$\backslash$\\
  10-20\%&139.4$\pm$3.2&270.0$\pm$9.0&1&1636$\pm$542&$\backslash$\\
  20-40\%&138.2$\pm$2.9&267.1$\pm$8.4&1&999$\pm$294&$\backslash$\\
  \midrule
  \multicolumn{6}{c}{$\snn=$ 7.7 GeV Pariticle Set I} \\
  0-10\% &143.3$\pm$2.1&412.5$\pm$14.1&1.15$\pm$0.06&687$\pm$100&4.7/6\\
  10-20\%&143.4$\pm$2.0&400.9$\pm$13.0&1.04$\pm$0.05&507$\pm$71&3.4/6\\
  20-40\%&145.0$\pm$2.2&387.5$\pm$13.7&0.92$\pm$0.04&293$\pm$43&3.8/6\\
  40-80\%&146.9$\pm$2.3&361.0$\pm$13.2&0.70$\pm$0.03&79.9$\pm$11.8&7.1/6\\
  \midrule
  \multicolumn{6}{c}{$\snn=$ 7.7 GeV Pariticle Set II} \\
  0-10\% &139.9$\pm$1.2&388.9$\pm$4.7&1.15$\pm$0.06&871$\pm$80&15.1/8\\
  10-20\%&140.6$\pm$1.1&382.6$\pm$4.4&1.05$\pm$0.05&616$\pm$54&10.8/8\\
  20-40\%&141.5$\pm$1.1&366.3$\pm$3.7&0.93$\pm$0.04&370$\pm$31&12.0/8\\
  40-80\%&143.3$\pm$1.2&338.8$\pm$3.8&0.71$\pm$0.03&99.7$\pm$8.8&11.9/8\\
  \bottomrule
  \end{tabular*}
\end{table}

\begin{figure}[!htb]
  \includegraphics[width=\hsize]{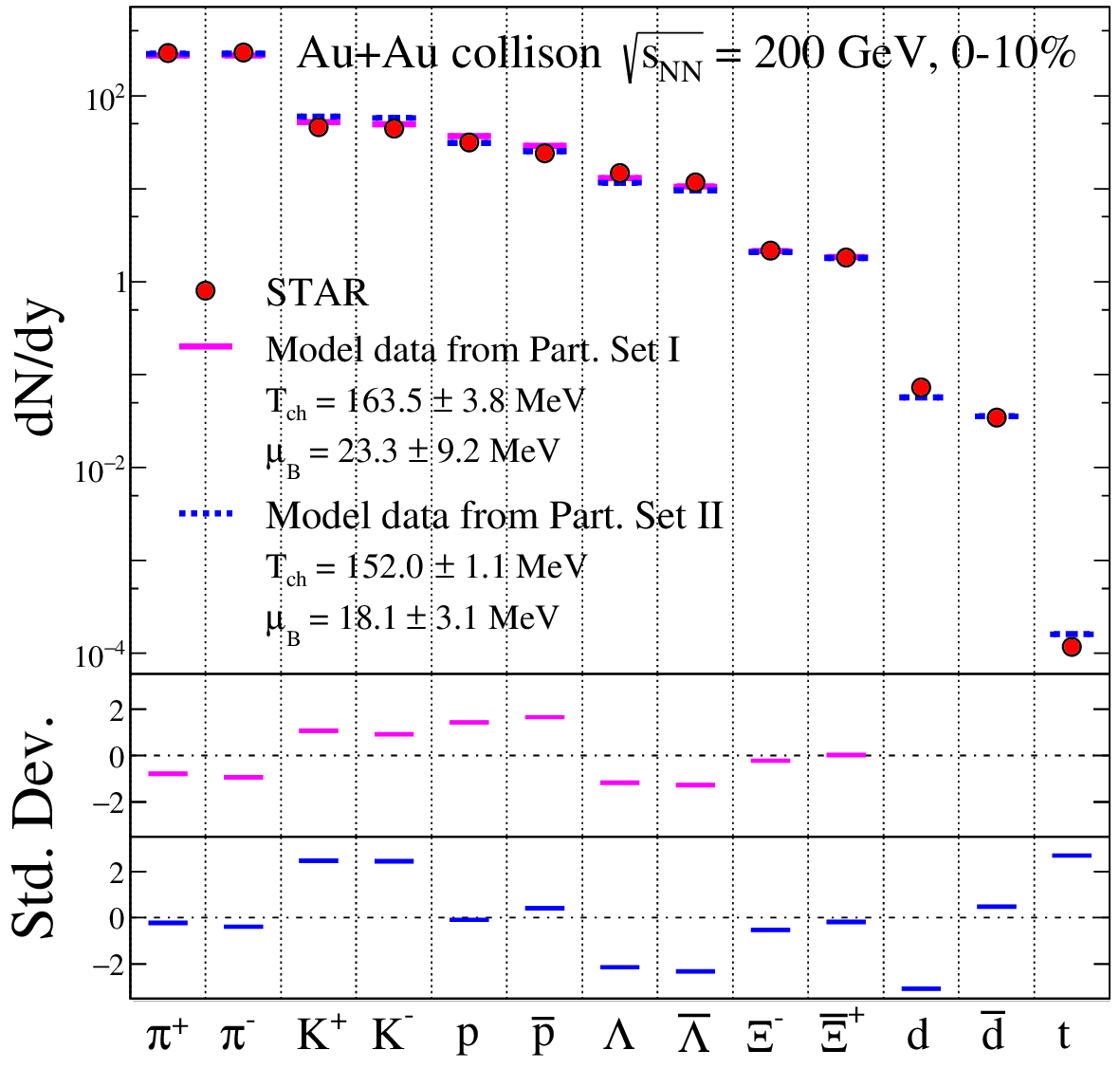}
  \caption{Particle yield and predictions of the Thermal-FIST package. Upper panel: $dN/dy$ values (red circles) from different hadrons and nuclei, measured at mid-rapidity, are compared with the fit results from particle set I (magenta solid bars) and II (blue dashed bars). The experimental data are from the STAR Collaboration for $0-10\%$ Au+Au collisions at $\snn=200$ GeV. Lower panel shows the standard deviations $\sigma$ between the fit and experimental data.}
  \label{fig1} 
\end{figure}

The upper panel of Figure~\ref{fig1} presents the comparison between particle yield per unit rapidity $dN/dy$ and the Thermal-FIST package fitting results in Au+Au collision at $\snn=$200 GeV within the 0-10\% centrality interval. The STAR experimental data is depicted with red circles, while the model data is represented by bars of different colors. We obtain different freeze-out parameters from these two particle sets. When light nuclei are not considered (magenta solid bars), $T_{\text{ch}}$ is $163.5\pm3.8$ MeV, which significantly decreases to $152.0\pm1.1$ MeV after including light nuclei yield in the fitting (blue dashed bars). The inclusion of light nuclei results in a noticeably lower fitted temperature. Additionally, as shown in Table~\ref{tab2}, if only light nuclei yield are used for fitting, this temperature further reduces to $137.6\pm3.4$ MeV. Similar to the conclusions in reference~\cite{RN308}, we can speculate that not only do light flavor and strange flavor have different freeze-out temperatures, but light nuclei also achieve a lower freeze-out temperature. The three experience production at different freeze-out surfaces. A possible reason is that the production of light nuclei occurs at the later stage of the collision, at least within the STAR energy region, where the production mechanism of light nuclei is through coalescence. Since in the coalescence model, light nuclei originate from the recombination of nucleons, we have also investigated the effect of including the yields of $p(\bar{p})$ in the fitting for Particle Set III. We found that the chemical freeze-out temperature $T_{\text{ch}}$ increased by $1\sim2$ MeV. For instance, in the 0-10\% central collisions at 200 GeV, the inclusion of the yield $p(\bar{p})$ resulted in a chemical freeze-out temperature that changed from $137.6\pm3.4$ MeV to $138.4\pm1.9$ MeV. Although there is an increase, this value remains significantly lower than the chemical freeze-out temperature obtained from Particle Set II.

The $\mu_B$ obtained from the two fittings remain consistent within error. The lower panel of Figure~\ref{fig1} shows the standard deviation (Std. Dev.) of different particles in the two fittings. It is evident that the deviation of strange flavor particles $K^{\pm}$, $\Lambda(\bar{\Lambda})$, and $\Xi^{\pm}$ significantly increases after the inclusion of light nuclei, while the deviation for light flavor particles $\pi^{\pm}$, and $p(\bar{p})$ decreases. This can also be simply explained by our previous conclusions: since the freeze-out temperature difference between strange flavor particles and light nuclei is significant, and the difference between light flavor particles and light nuclei is relatively smaller, the inclusion of light nuclei cannot fit the strange flavor particles well but can improve the fitting for light flavor particles.

\begin{figure}[!htb]
  \includegraphics[width=\hsize]{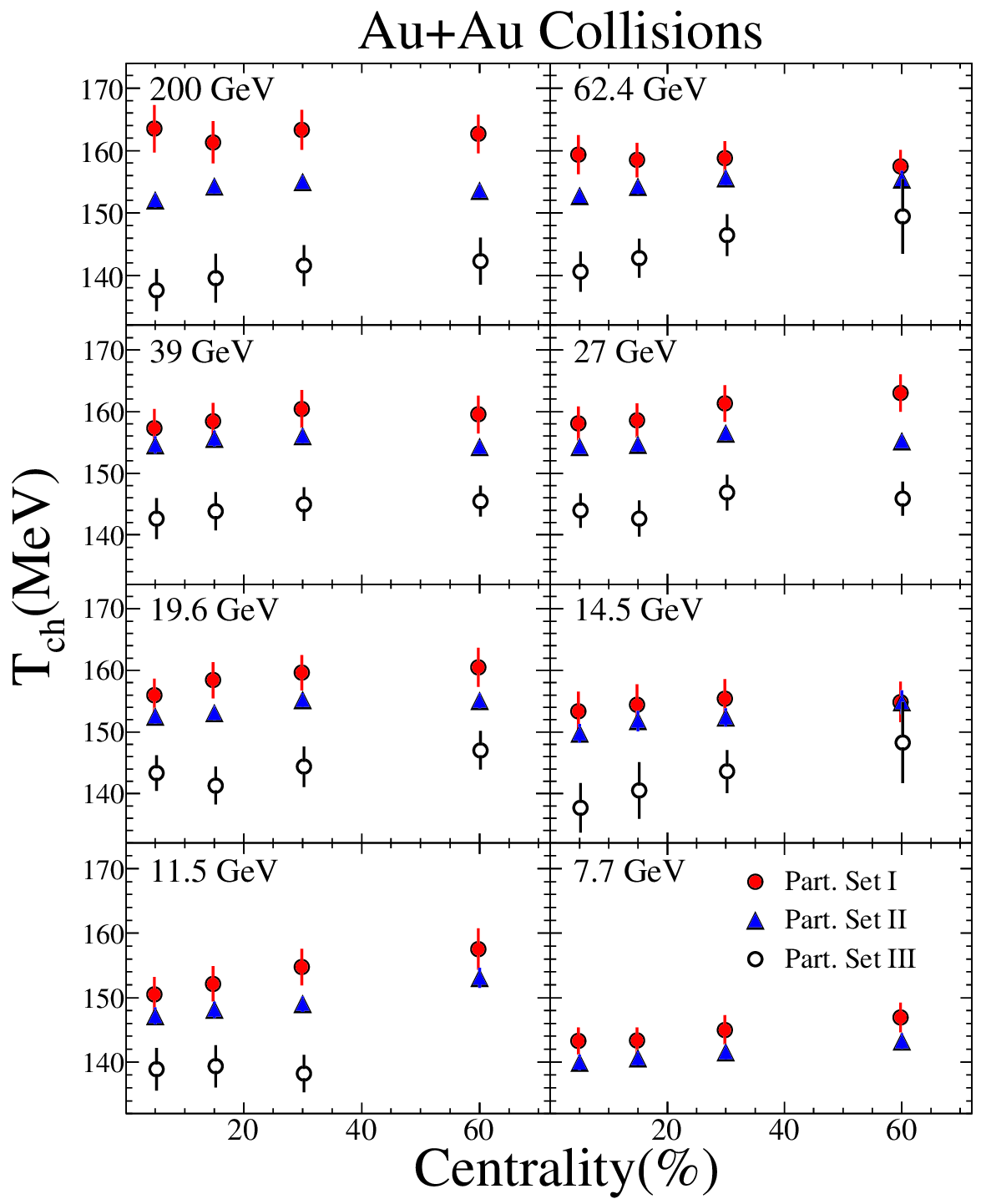}
  \caption{Centrality dependence of chemical freeze-out parameters $T_{\textrm{ch}}$ from the Thermal-FIST fit with Particle set I(solid red circles), Particle set II(solid blue triangles), and Particle set III(open black circles) listed in Table~\ref{tab2} in Au+Au collisions at $\snn =$ 7.7-200 GeV.}
  \label{fig2} 
\end{figure}

Figure~\ref{fig2} presents a series of graphs displaying the centrality dependence of the chemical freeze-out temperature $T_{\textrm{ch}}$ from the Thermal-FIST fit with three different Particle sets for in Au+Au collisions at $\snn =$ 7.7-200 GeV. It can be observed that with the increase in centrality, that is, from central to peripheral collisions, for different collision energies and different particle sets, $T_{\textrm{ch}}$ exhibits a consistent trend, remaining independent of centrality within error. Additionally, it is evident that the three particles sets of yield three distinct $T_{\textrm{ch}}$ range in the fitting. When light nuclei are included in the fitting, as seen in the transition from Particle set I to Particle set II (solid circles to solid triangles in the figure), $T_{\textrm{ch}}$ decreases. For instance, at $\snn=$200 GeV, the decrease is about 10 MeV, and at other collision energies, it is roughly 5 MeV. More notably, when considering only light nuclei, as in Particle set III (opened circles in the figure), $T_{\textrm{ch}}$ drops even further compared to Particle set II, which include all particles. For example, at $\snn=$200 GeV, this difference reaches about 14 MeV, and at the lower energy of 11.5 GeV, it is around 9 MeV. These differences clearly separating the opened circles from the other two categories in the figure. This also indicates that light nuclei indeed reach chemical freeze-out at a different time compared to other particles, being produced later in the collision process.

\begin{figure}[!htb]
  \includegraphics[width=\hsize]{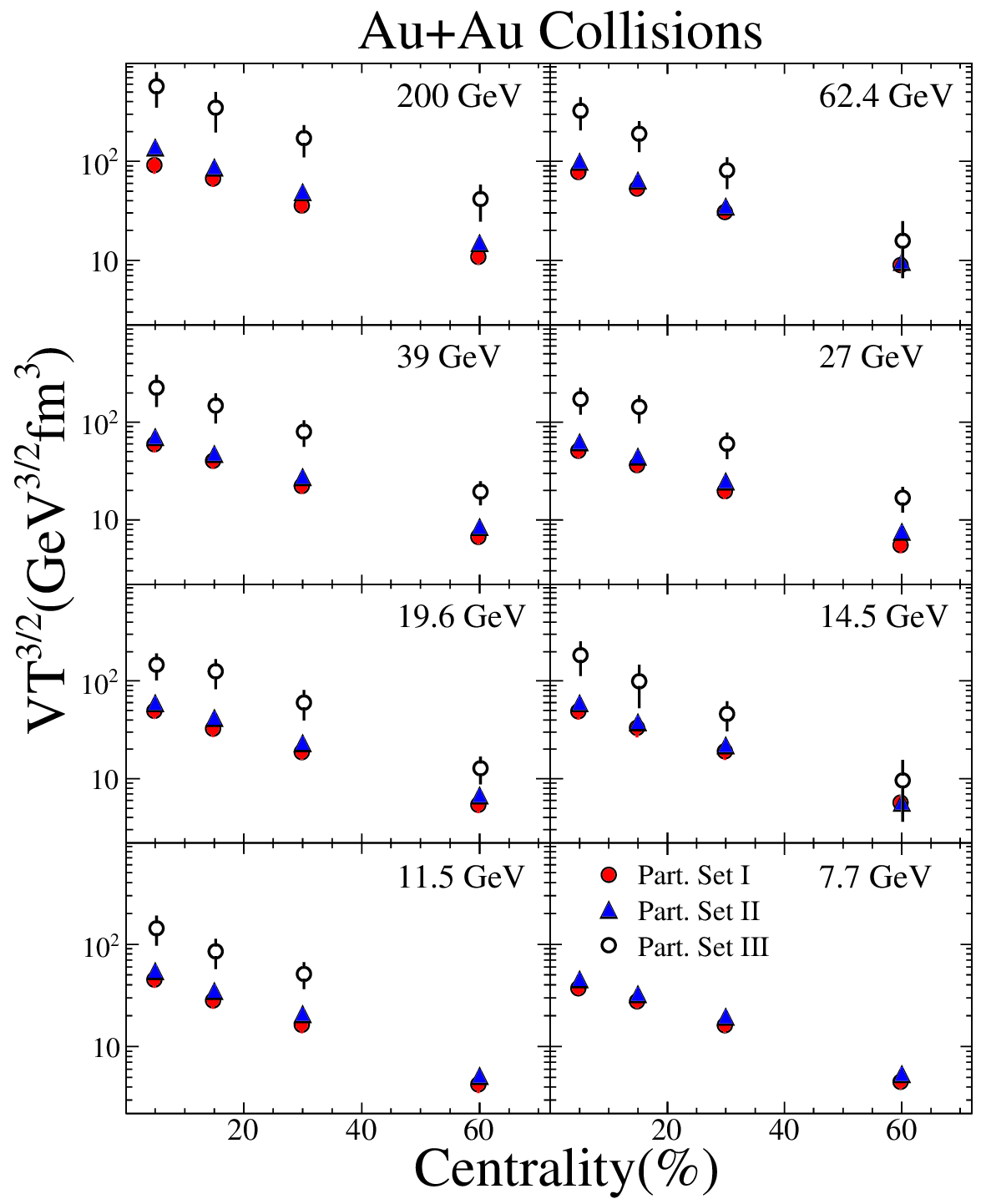}
  \caption{Centrality dependence of $VT^{3/2}$ calculated by the chemical freeze-out parameters from the Thermal-FIST fit with Particle set I(solid red circles), Particle set II(solid blue triangles), and Particle set III(open black circles) listed in Table~\ref{tab2} in Au+Au collisions at $\snn =$ 7.7-200 GeV.}
  \label{fig3} 
\end{figure}

To systematically study the stages of particle production, we present in Figure~\ref{fig3} the variation of $VT^{3/2}$ with collision centrality for different particle sets at various collision energies. According to the Sackur-Tetrode equation~\cite{grimus}, under the non-relativistic $VT^{3/2}$ is directly related to the entropy per nucleon ($S/N$)~\cite{RN165}. 
\begin{equation}
  \frac{S}{N}=\ln\frac{VT^{3/2}}{N} + \textrm{const.}
\end{equation}
After chemical freeze-out, the number of particles remains essentially constant, so the magnitude of $VT^{3/2}$ can essentially reflect the change in the system's entropy. It is well known that the evolution of an isolated system always proceeds in the direction of increasing entropy. In other words, the magnitude of $VT^{3/2}$ can reflect the sequence of the system's evolution process; that is, for the same system, a smaller $VT^{3/2}$ indicates an earlier stage of system evolution, while a larger $VT^{3/2}$ indicates a later stage  of system evolution. It is evident from Figure~\ref{fig3} that when the yield of light nuclei are included, that is, transitioning from particle set I to particle set II, the system's $VT^{3/2}$ increases. If only the yield of light nuclei are considered, an even larger $VT^{3/2}$ is obtained, which also indicates that at RHIC energies, light nuclei are likely produced later in the evolution, rather than the slightly earlier time of strange hadron freeze-out. Similarly, it can be observed that for a certain collision energy and a certain particle set, the $VT^{3/2}$ is larger in the central collisions, meaning that each particle carries more entropy and there is a higher degree of disorder in central collisions. Additionally, for a certain collision centrality and a certain particle set, $VT^{3/2}$ decreases with the decreasing of collision energy, meaning that the entropy carried by each particle is smaller at lower energies compared to higher energies, which is quite understandable.

\begin{figure}[!htb]
  \includegraphics[width=\hsize]{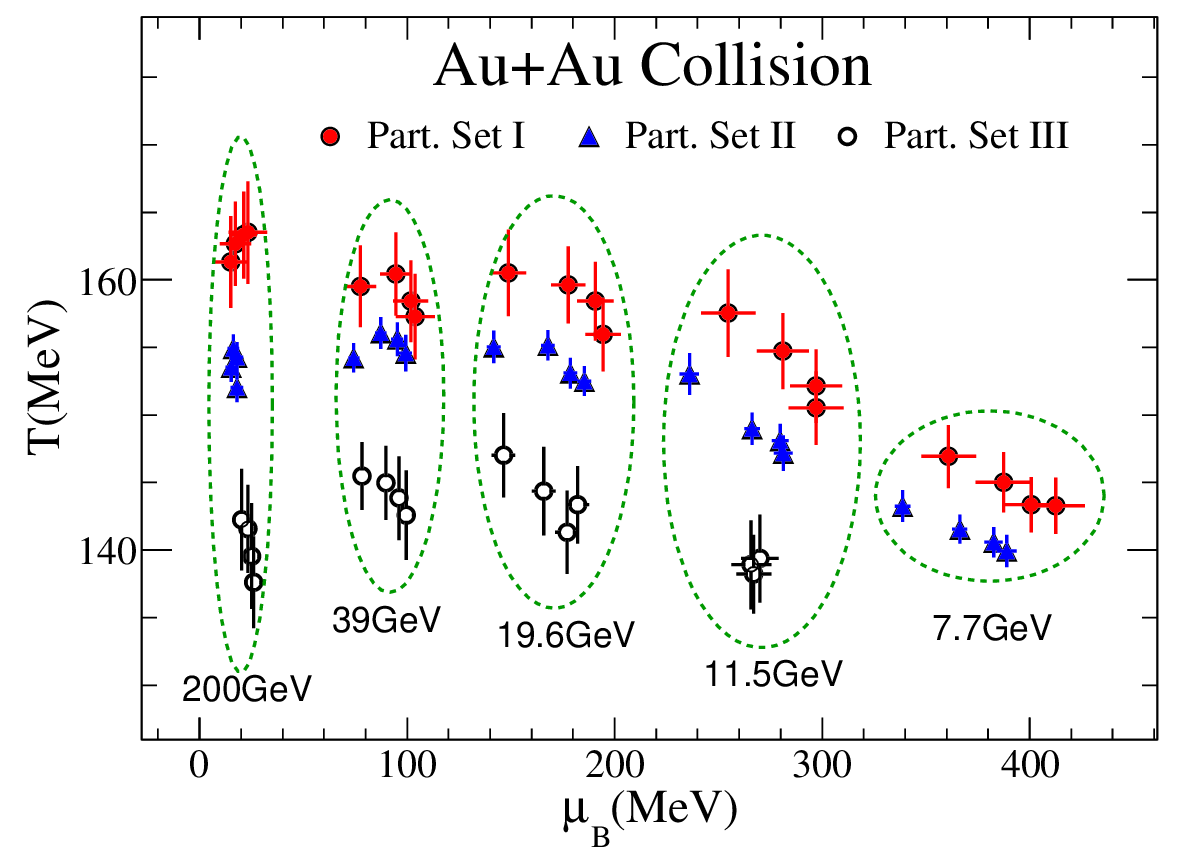}
  \caption{The chemical freeze-out parameters $T_{\textrm{ch}}$ and $\mu_B$ as a function of collision energy in Au+Au collisions. Data point represent the results from the Thermal-FIST fit with Particle set I(solid red circles), Particle set II(solid blue triangles), and Particle set III(open black circles) listed in Table~\ref{tab2} in Au+Au collisions at $\snn =$200, 39, 19.6, 11.5, and 7.7GeV. Elliptical dashed lines represent the results corresponding to different collision energies.}
  \label{fig4}
\end{figure}

To provide a clear visual representation of the system's location on the two-dimensional 
$T-\mu_B$ phase diagram at chemical freeze-out, Figure~\ref{fig4} presents the chemical freeze-out parameters, $T_{\textrm{ch}}$ and $\mu_B$, as a function of collision energy in Au+Au collisions. To enhance readability and mitigate the visual clutter that can result from overlapping data points, the results for collision energies at 62.4 GeV, 27 GeV, and 14.5 GeV have been intentionally excluded. Additionally, the outcomes for different collision energies are delineated using elliptical dashed lines. It can be observed that within the same particle set, $T_{\textrm{ch}}$ decreases with decreasing collision energy. However, it is interesting to note that under Particle Set III, which includes only light nuclei yields, this dependence on collision energy is quite weak, remaining around approximately $T_{\textrm{ch}}\approx$140 MeV. In other words, the chemical freeze-out temperature for light nuclei, or their coalescence temperature, is roughly 140 MeV. This observation may potentially explain why the freeze-out temperatures derived from the top energy at RHIC~\cite{RN24} are higher than those obtained from LHC results~\cite{RN102}, as the LHC fits incorporated multiple light nuclei yields. It is observed that as the collision energy decreases,$\mu_B$ increases. This is attributed to a reduction in the production of antibaryons, leading to an increased net baryon density and consequently, a rise in $\mu_B$. Moreover, it can be seen that different particle sets yield rather consistent results for $\mu_B$. Furthermore, the three distinct particle sets clearly define three separate bands on the diagram, indicating that at RHIC energies, light nuclei and other hadrons likely exist on distinct freeze-out hyper-surfaces.

\begin{figure}[!htb]
  \includegraphics[width=\hsize]{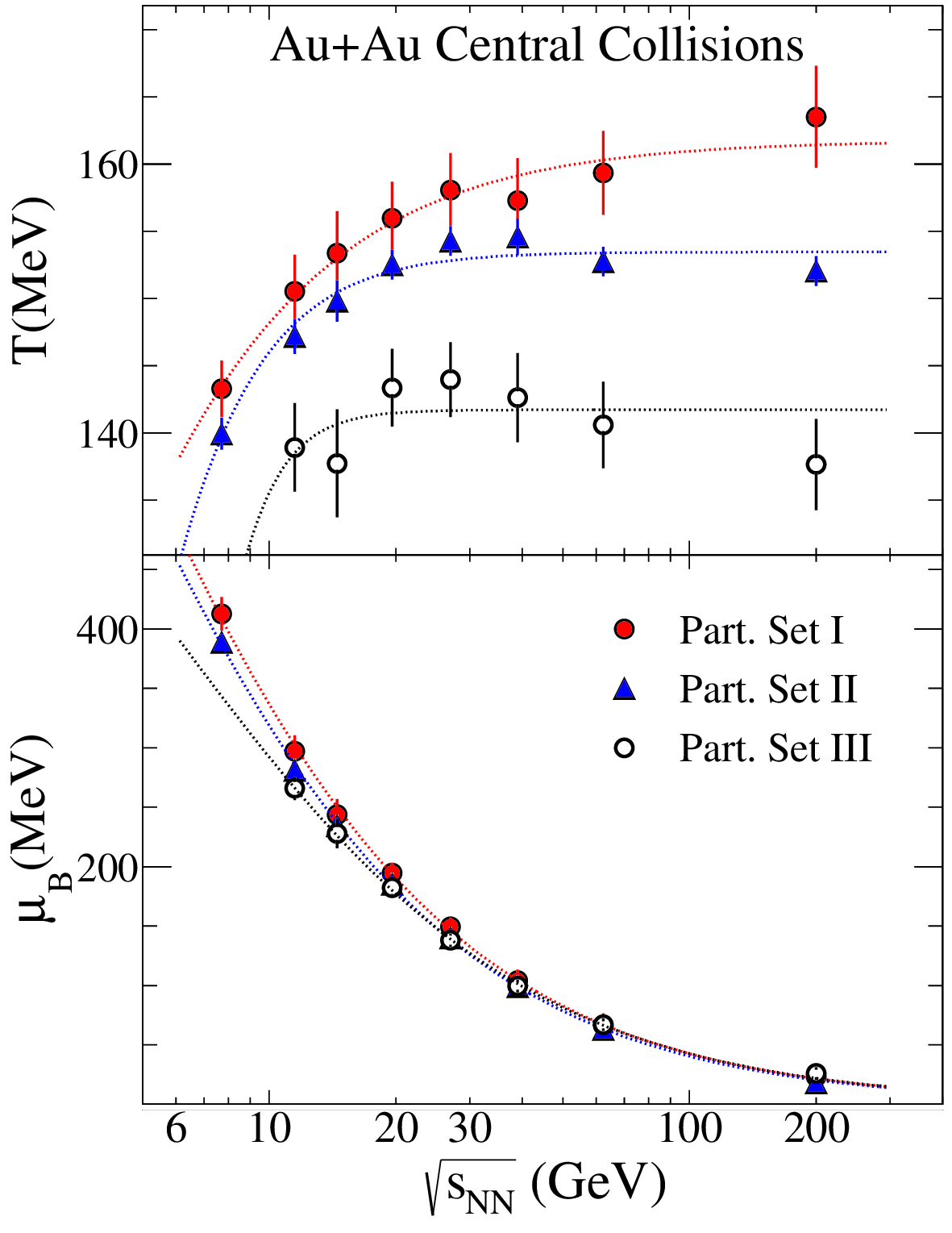}
  \caption{Chemical freeze-out parameters $T_{\textrm{ch}}$ and $\mu_B$ as a function of the center-of-mass collision energy for Au+Au central collisions. The curves are parametrization for $T_{\textrm{ch}}$ and $\mu_B$ from Eq.~(\ref{eqtmub}).}
  \label{fig5}
\end{figure}

Formulas for the parameterization of the chemical freezing parameter $T_{\textrm{ch}}$ and $\mu_B$, using collision energy $\snn$ as a variable, are detailed in reference~\cite{RN102}. The expressions are as follows:
\begin{equation}\label{eqtmub}
  \begin{split}
  T_{\textrm{ch}} &= T_{\textrm{ch}}^{\textrm{lim}}\frac{1}{1+\exp\{a-\ln[\snn(\textrm{GeV})]/b\}}\\
  \mu_B &= \frac{c}{1+d\snn(\textrm{GeV})}
  \end{split}
\end{equation}
These parameterized equations utilize experimental fit data both from STAR and LHC collaborations. The chemical freeze-out parameters for these experiments were fitted using various particle sets, as previously discussed. It is evident from the previous discussion that incorporating light nuclei in the fitting process yields significantly influences the derived chemical freeze-out temperatures. Thus, parameterizing the chemical freeze-out parameters from different particle sets is crucial. Figure~\ref{fig5} presents the parameterized chemical freeze-out parameters, including $T_{\textrm{ch}}$ and $\mu_B$, obtained through fits with various particle sets in central Au+Au collisions. The detailed parameters are provided in Table~\ref{tab4}. It is evident that the fitted parameters exhibit notable differences, particularly $T_{\textrm{ch}}^{\textrm{lim}}$, which signifies the highest temperature reachable during hadronic chemical freeze-out. This parameter also represents the maximum temperature at which the hadron resonance gas can exist before transitioning to a Quark-Gluon Plasma. For fits excluding light nuclei yields, such as in particle set I, this value is 161.8$\pm$3.6 MeV, which is higher than the value reported in reference~\cite{RN102}. When light nuclei yields are incorporated into the fitting process, as in particle set II, the parameterized $T_{\textrm{ch}}^{\textrm{lim}}$ decreases to 153.5$\pm$1.4 MeV. Similarly, we observe that other parameters align with those obtained in reference ~\cite{RN102}, indicating that the results from particle set II fit are consistent with the parametrized equations derived therein. However, when considering only light nuclei yields, as in particle set III, $T_{\textrm{ch}}^{\textrm{lim}}$ further reduces to 141.7$\pm$1.4 MeV, which is notably lower than in the previous two scenarios. This suggests that light nuclei and other hadrons likely exist on two distinct freeze-out hyper-surfaces. Due to the absence of fit results for 7.7 GeV in particle set III, we are unable to determine the behavior of $T_{\textrm{ch}}$ at lower energies from the existing data. Additional experimental data encompassing a variety of light nuclei yields could enhance characterization of this energy dependency.

\begin{table}[!htb]
  \caption{\label{tab4}The parameters in energy-dependent $T_{\textrm{ch}}$ and $\mu_B$ in central collisions.}
  \begin{tabular*}{\hsize}{@{\extracolsep{\fill}} cccccc}
  \toprule
  Particle Set	&	$T_{\textrm{ch}}^{\textrm{lim}}$(MeV) & $a$ & $b$ & $c$(MeV) & $d$\\
  \midrule
  Ref.~\cite{RN102} & 158.4$\pm$1.4 & 2.60 & 0.45 & 1307         & 0.286\\
  I	                &	161.8$\pm$3.6 & 0.52 & 0.79 & 1442.2$\pm$306.5 & 0.329\\
  II	              &	153.5$\pm$0.6 & 2.82 & 0.40 & 1356.8$\pm$104.3  & 0.326\\
  III	              &	141.7$\pm$1.4 & 8.00 & 0.21 & 829.7$\pm$127.4  & 0.184\\
  \bottomrule
  \end{tabular*}
\end{table}

\begin{figure}[!htb]
  \includegraphics[width=\hsize]{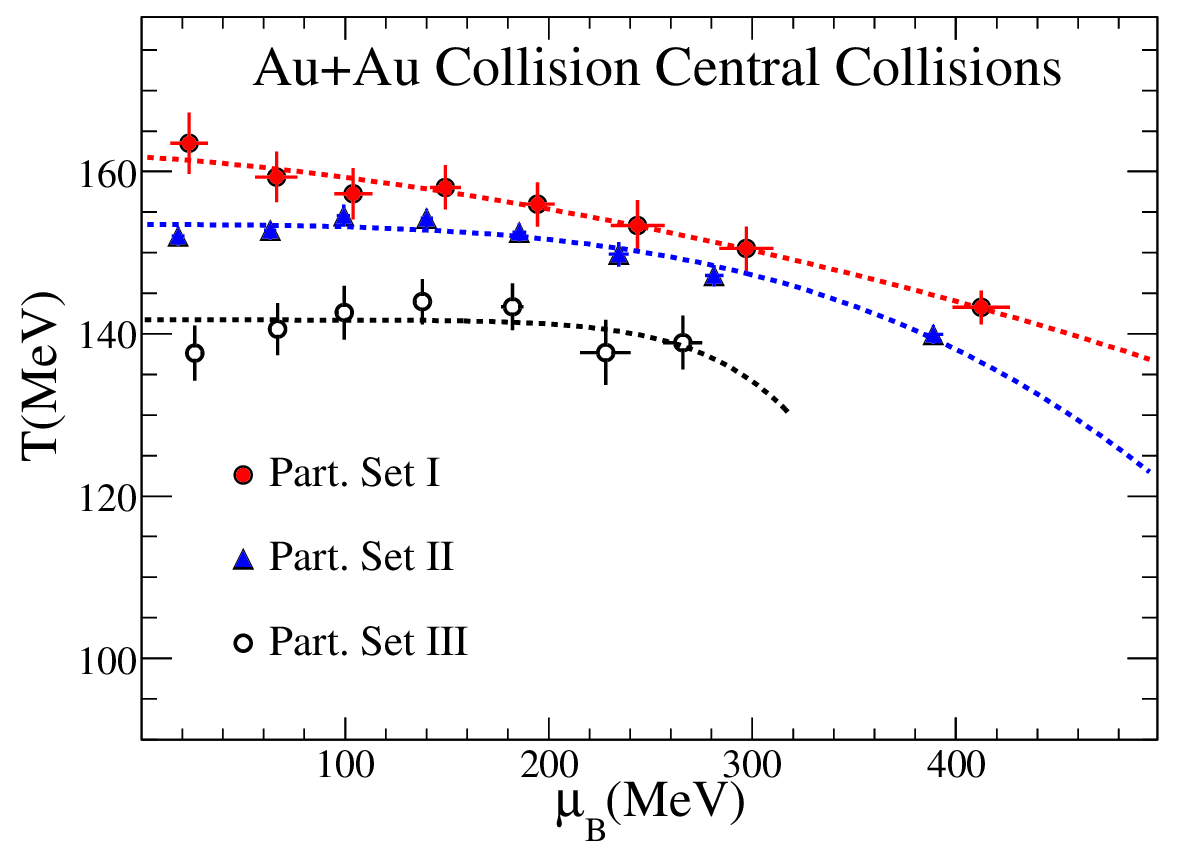}
  \caption{Chemical freeze-out parameters temperature $T_{\textrm{ch}}$ versus baryon chemical potential $\mu_B$ from central Au+Au collisions at $\snn=$7.7-200 GeV. Data point represent the results from the Thermal-FIST fit with Particle set I(solid red circles), Particle set II(solid blue triangles), and Particle set III(open black circles) listed in Table~\ref{tab2}. The curves are parametrization for $T_{\textrm{ch}}$ and $\mu_B$ from Eq.~(\ref{eqtmub}).}
  \label{fig6}
\end{figure}

Through the energy-dependent relationship of the chemical freeze-out parameters $T_{\textrm{ch}}$ and $\mu_B$, we are not only able to calculate the yields of various hadrons but also plot the chemical freeze-out line on a two-dimensional phase diagram at the RHIC energies. This line can provide information for determining the phase boundary from hadronic matter to Quark-Gluon Plasma. In Figure~\ref{fig6}, we present this chemical freeze-out line. Due to insufficient light nuclei yield data at 7.7 GeV, for Particle Set III, we only depict the region where $\mu_B$< 320 MeV. It is evident that the chemical freeze-out lines differ for various particle sets. The line for light nuclei (particle set III) and hadrons without light nuclei (particle I) are distinctly separated, indicating that light nuclei and other hadrons freeze-out at different hyper-surfaces. In conjunction with the results from Ref.~\cite{RN308}, we can identify at least three different chemical freeze-out temperatures at the RHIC energies: a light flavor freeze-out temperature $T_L$=150.2$\pm$6 MeV, a strange flavor freeze-out temperature $T_s$=165.1$\pm$2.7 MeV, and a light nuclei freeze-out temperature $T_{\textrm{ln}}$=141.7$\pm$1.4 MeV. The fireball produced after relativistic heavy-ion collisions exhibits three distinct freeze-out hyper-surfaces. The first pertains to strange hadrons, which are expected to have relatively smaller hadronic interaction cross sections compared to light hadrons. Consequently, the primary yield of strangeness particles will experience little change from the stage of phase transition due to hadronic interactions~\cite{RN90,RN147}. As a result, they can carry the temperature information from the quarks' chemical equilibrium, which is higher than the temperature at which light hadron chemical freeze-out occurs, and exhibit a clear sequence of chemical freeze-out throughout the system's evolution. For light nuclei, their yields evolve as the system progresses, indicating that they reach chemical equilibrium much later, i.e., they freeze out of the system at lower temperatures. This observation is crucial for understanding the complex dynamics of particle production and the thermodynamic evolution of the medium created in heavy-ion collisions.

It should be emphasized that our conclusions are based on the existing experimental data from the RHIC energies, meaning that our findings are at least credible at the RHIC energies. To verify whether our conclusions hold true in the LHC energy, we repeated the fitting process using experimental data from the LHC. During the fitting process, we initially replicated the particle set from Ref.~\cite{RN102}, and subsequently performed fits using the particle sets I, II, and III as presented in this study. It is important to note that, due to the absence of triton experimental data at the LHC, we utilized the yield data of $^3\textrm{He}$ to proxy for the yield of tritons in the fitting procedure. We also utilized all available light nuclei yield data from the LHC, including $d, \bar{d}, ^3\textrm{He}, ^3\bar{\textrm{He}}, ^4\textrm{He},$ and, $^4\bar{\textrm{He}}$, for the fitting, which is label as Particle Set IV. The final set, label as Particle Set V, included the yields of hypertriton $^3_{\Lambda}\textrm{H}$ and its antiparticle $^3_{\bar{\Lambda}}\bar{\textrm{H}}$. A total of six particle sets were fitted, and the results are listed in Table~\ref{tab5}.

\begin{table}[!htb]
  \caption{The Thermal-FIST Grand Canonical Ensemble fit for 0-10\% Pb+Pb collision at $\snn=$2.76 TeV. For the fits in particle set I, II, and V, $T_{\text{ch}}$, $\mu_B$, $\gamma_s$, and $V$ were used as free parameters. For the fit in particle set III and IV, $\gamma_s$ were fixed to unity.}
  \label{tab5}
  \footnotesize
  \begin{tabular*}{\hsize} {@{\extracolsep{\fill} } cccccc}
  \toprule
  Particle Set&$T_{\text{ch}}$(\si{MeV}) & $\mu_B$(\si{MeV}) & $\gamma_s$ & $V$(\si{fm^3}) & $\chi^2$/dof \\
  \midrule
  Ref.~\cite{RN102} &154.5$\pm$1.4&1.7$\pm$4.4 &1.11$\pm$0.03&4200$\pm$383 &30.1/18\\
  I                 &150.6$\pm$2.2&1.5$\pm$6.2 &1.14$\pm$0.04&5166$\pm$641 &21.5/6\\
  II                &152.2$\pm$1.8&3.2$\pm$5.1 &1.13$\pm$0.04&4794$\pm$530 &23.0/9\\
  III               &157.8$\pm$8.3&0.8$\pm$10.5&1            &3037$\pm$2293&$\backslash$\\
  IV                &159.2$\pm$1.3&1.3$\pm$6.6 &1            &2621$\pm$1331&1.2/3\\
  V                 &159.2$\pm$4.9&1.7$\pm$5.8 &1.35$\pm$0.35&2623$\pm$1329&1.2/4\\
  \bottomrule
  \end{tabular*}
\end{table}

As observed in the table, in the LHC energy, specifically for Pb+Pb collisions at $\snn=$2.76 TeV, the inclusion or exclusion of light nuclei yields in the fitting process does not significantly affect the derived chemical freeze-out temperature $T_{\text{ch}}$. Fitting using the yield of all particles (first row in the table), fitting without the yield of light nuclei (Particle Set I), and fitting with the inclusion of $d$, $\bar{d}$, and $^3\textrm{He}$ yields (Particle Set II) result in chemical freeze-out temperatures $T_{\text{ch}}$ that are consistent within error. Even other chemical freeze-out parameters such as $\mu_B$, $\gamma_s$, and $V$ are found to be consistent within the error. The primary difference is that fitting with more particles reduces the $\chi^2$/dof. More intriguingly, when only the yields of light nuclei are incorporated into the fitting process, a higher $T_{\text{ch}}$ is obtained, around $158-159$ MeV. Regardless of the light nuclei particle sets, the resulting $T_{\text{ch}}$ is consistent, and this temperature is higher than that of other particles. This is in contrast to the conclusions drawn from the RHIC energies, where the chemical freeze-out temperature of light nuclei is approximately 10 MeV lower than that of light flavor hadrons, whereas at the LHC $\snn=$2.76 TeV, the chemical freeze-out temperature of light nuclei is about 10 MeV higher than that of light flavor hadrons. Given the difference in experimental data errors between RHIC and LHC, we investigated the impact of these data errors on the fitting results by artificially reducing the original experimental errors by 50\% for the 0-10\% central Au+Au collisions at $\sqrt{s_{\textrm{NN}}} = 200$  GeV. The result of the fitting indicated that this adjustment did not alter the central values of the fitted parameters but rather affected the magnitude of the errors. For instance, with Particle Set I,  chemical freeze-out temperature shifted from 163.5$\pm$3.8 MeV to 163.5$\pm$1.9 MeV. 

Recently, the ALICE Collaboration published thermodynamic fitting results for Pb+Pb collisions at 5.02 TeV, where they found that a chemical freeze-out temperature of $155\pm2$ MeV provided an excellent fit to the experimental data at $\sqrt{s_{\textrm{NN}}} = 5.02$ TeV~\cite{RN318}. This result is similar to what we obtained using Particle Set II at 2.76 TeV. Additionally, the ALICE Collaboration has published results on light nuclei yields~\cite{RN316}. To verify our conclusions at 5.02 TeV, we performed a similar fitting using these light nuclei yields, selecting yields of four particles: $d, \bar{d}, ^3\textrm{He}, ^3\bar{\textrm{He}}$. We obtained a chemical freeze-out temperature of $145.8\pm0.7$ MeV listed in Table~\ref{tab6}, which is lower than the result reported in the ALICE paper. Based on the current results, we find that the ALICE Pb+Pb collisions at $\sqrt{s_{\textrm{NN}}} = 2.76$ TeV and 5.02 TeV yield opposite outcomes. At the LHC $\sqrt{s_{\textrm{NN}}} = 2.76$ TeV, the chemical freeze-out temperature of light nuclei is higher than that of light flavor and strange flavor hadrons. However, at the ALICE Pb+Pb $\sqrt{s_{\textrm{NN}}} = 5.02$ TeV, the chemical freeze-out temperature of light nuclei is lower than that of light flavor and strange flavor hadrons. This result is consistent with our conclusions at RHIC and also aligns with the findings in Ref.~\cite{RN319}, which state that deuterons and tritons chemically freeze out to the late hadronic stage. 

\begin{table}[!htb]
  \caption{The Thermal-FIST Grand Canonical Ensemble fit for 0-10\% Pb+Pb collision at $\snn=$5.02 TeV. For the fit in particle set III, $\gamma_s$ were fixed to unity.}
  \label{tab6}
  \footnotesize
  \begin{tabular*}{\hsize} {@{\extracolsep{\fill} } cccccc}
  \toprule
  Particle Set&$T_{\text{ch}}$(\si{MeV}) & $\mu_B$(\si{MeV}) & $\gamma_s$ & $V$(\si{fm^3}) & $\chi^2$/dof \\
  \midrule
  Ref.~\cite{RN318} &155$\pm$2&0.73$\pm$0.52 &  &  &0.85/4\\
  III               &145.2$\pm$1.5&0.9$\pm$2.3&1            &12782$\pm$2289&$\backslash$\\
  III+$^3\bar{\textrm{He}}$&145.8$\pm$0.7&0.0$\pm$0.5&1            &12033$\pm$1139&0.15/1\\
  \bottomrule
  \end{tabular*}
\end{table}

The peculiarity of the results at 2.76 TeV might arise from two main considerations. Firstly, the production mechanisms of light nuclei, may differ at 2.76 TeV. However, we do not have compelling reasons to suspect that the particle production mechanisms vary among the top energy at RHIC,  2.76 TeV and 5.02 TeV collisions at the LHC, all of which are high-energy collision systems. Secondly, there may be energy-dependent effects that influence the dynamics of the collisions and the resulting particle yields. Yet, based on the current findings, we cannot provide a definitive answer.This hypothesis necessitates further investigation to gain a deeper understanding of the underlying processes for the production of light nuclei.

\section{Summary}
Thermodynamic statistical model fitting enables precise determination of chemical freeze-out parameters $T_{\text{ch}}$ and $\mu_B$ during relativistic heavy-ion collisions. These parameters offer crucial insights into the phase transition from QGP to hadronic matter. In this study, we utilized the Thermal-FIST Grand Canonical Ensemble fit to analyze the experimental data from RHIC Au+Au collisions at collision energies of $\snn=$ 7.7--200 GeV. The chemical freeze-out parameters of the system were determined using various particle sets and collision centralities. Different particle sets yield different chemical freeze-out temperatures, with light nuclei achieving lower freeze-out temperatures. Additionally, we examined $VT^{3/2}$ for different particle sets, which reflected the entropy per particle ($S/N$) of the system. We found that when considering only light nuclei, $VT^{3/2}$ was larger, indicating higher entropy and suggesting that light nuclei are produced at a later system evolution stage. We also obtained the parameterized chemical freeze-out parameters $T_{\text{ch}}$ and $\mu_B$. Similar to the conclusions of previous studies, these findings indicate the existence of multiple chemical freeze-out hypersurfaces for different hadron types at RHIC energies. Concurrently, the experimental data indicate three distinct chemical freeze-out temperatures: a light-flavor freeze-out temperature $T_L$ = 150.2$\pm$6 MeV, a strange-flavor freeze-out temperature $T_s$ = 165.1$\pm$2.7 MeV, and a light-nuclei freeze-out temperature $T_{\textrm{ln}}$ = 141.7$\pm$1.4 MeV. When applying our research methodology to LHC energies, we observed similar results for Pb+Pb 5.02 TeV, where the light-nuclei chemical freeze-out temperature was lower than that for light-flavor hadrons. However, at Pb+Pb 2.76 TeV, we did not observe a lower light-nuclei freeze-out temperature. By contrast, the light-nuclei chemical freeze-out temperature was approximately 10 MeV higher than that for light-flavor hadrons. This indicates that energy-dependent effects influence the collision dynamics and the resulting particle yields. We aim to address this issue in future work.

\bibliography{CFO}

\end{document}